\documentclass[10pt]{article}
\usepackage{fullpage,amssymb,amsmath}
\usepackage[dvips]{epsfig}

\def\##1{{\bf #1}}
\def\=#1{\underline{\underline{#1}}}

\def\+#1{\underline{\bf #1}}
\def\*#1{\underline{\underline{\bf #1}}}

\def\r#1{(\ref{#1})}
\def\l#1{\label{#1}}
\def\c#1{\cite{#1}}

\def\le{\left(}
\def\ri{\right)}
\def\les{\left[}
\def\ris{\right]}
\def\lec{\left\{}
\def\ric{\right\}}

\def\.{\mbox{ \tiny{$^\bullet$} }}

\def\omegao{\omega_{\scriptscriptstyle 0}}

\def\eps{\epsilon}

\begin{document}

\LARGE
\begin{center}
{\bf On the application of  homogenization formalisms to active dielectric
composite materials}

\vspace{10mm} \large

 Tom G. Mackay\footnote{E--mail: T.Mackay@ed.ac.uk.}\\
{\em School of Mathematics and
   Maxwell Institute for Mathematical Sciences\\
University of Edinburgh, Edinburgh EH9 3JZ, UK}\\
\vspace{3mm}
 Akhlesh  Lakhtakia\footnote{E--mail: akhlesh@psu.edu}\\
 {\em NanoMM~---~Nanoengineered Metamaterials Group\\ Department of Engineering Science and Mechanics\\
Pennsylvania State University, University Park, PA 16802--6812, USA}

\end{center}

\vspace{4mm}

\normalsize

\begin{abstract}
The Maxwell Garnett and Bruggeman  formalisms were applied to
estimate the effective permittivity dyadic of active dielectric
composite materials. The active nature of the  homogenized composite
materials (HCMs)  arises from one of the component materials which
takes the form of InAs/GaAs quantum dots. Provided that the real
parts of the permittivities of the component materials have the same
sign, the Maxwell Garnett and Bruggeman formalisms give physically
plausible estimates of the HCM permittivity dyadic that are in close
agreement. However, if the real parts of the permittivities of the
component materials have different signs then there are substantial
differences between the Bruggeman and Maxwell Garnett estimates.
Furthermore, these differences becomes enormous~---~with the
Bruggeman estimate being physically implausible~---~as the imaginary
parts of the permittivities of the component materials tend to zero.

\end{abstract}


\section{Introduction}

Suppose that we have a mixture of two isotropic dielectric
materials, specified by the relative permittivity scalars $\eps_1$
and $\eps_2$.
Provided that the length scale of nonhomogeneities in
the mixture is small compared with electromagnetic wavelengths in
both component materials \c{M08}, the mixture can be treated as a
homogenized composite material (HCM).
 The familiar formalisms named in honour of Maxwell
Garnett \c{MG,Smith,Niklasson} and Bruggeman \c{Brugge,Goncharenko}
provide a means of estimating the relative permittivity of the
HCM.\footnote{These formalisms are readily extended to estimate the
relative permittivity of HCMs arising from a mixture of three or
more component materials \c{ML06}.} These formalisms are
well--established for dissipative HCMs arising from component
materials for which the real parts of $\eps_1$ and $\eps_2$ have the
same sign \c{L96}. However, we recently demonstrated that these
formalisms give rise to estimates of the HCM's relative permittivity
which are not physically plausible if the component materials are
weakly dissipative  and the real parts of $\eps_1$ and $\eps_2$ have
different signs \c{ML04,M07}. Furthermore, this limitation also
extends to the Bergman--Milton bounds on the HCM's relative
permittivity \c{DML07}.

In this short communication, we consider whether these limitations on
conventional homogenization formalisms also extend to active HCMs.
This topic is particularly timely, given the current huge interest
in metamaterials~---~many of which can be viewed as HCMs \c{M05}.
While certain metamaterials can be engineered to exhibit negative
refraction, the issue of dissipation presents a
major barrier to their practical implementation. In order to
overcome inherent dissipation, negatively refracting metamaterials containing active components
have been proposed recently \c{Noginov, Bratkovsky}.

In the following, double underlining signifies a 3$\times$3 dyadic,
with $\=0$ and $\=I$ denoting the null and identity dyadics
respectively.  The unit vector aligned with the Cartesian $z$ axis
is written as  $\hat{\#z}$. The real and imaginary parts of complex
quantities are identified by the prefixes `Re' and `Im',
respectively; and $i = \sqrt{-1}$.

\section{Investigation}

\subsection{Component materials}

We base our investigation on the metal--wire/quantum--dot
composite material proposed by Bratkovsky \emph{et al.} \c{Bratkovsky}.
That is, we consider a three--component composite material comprising silver
nanorods inserted into an array of InAs/GaAs quantum dots, with the
assembly being immersed in a host dielectric material. The silver
nanorods are aligned with the  Cartesian $z$ axis, while the quantum
dots are presumed to be spherical particles.
 The particles which comprise the host dielectric material
are assigned a spherical shape in the implementation of the
Bruggeman homogenization formalism. In contrast, no microstructure
  is assigned to the host dielectric medium in the implementation of
the Maxwell Garnett homogenization formalism. The volume fractions
of the silver nanorods, quantum dots and host dielectric material
are written as $f_{Ag}$, $f_{QD}$ and $f_h$, respectively, and we
have
\begin{equation}
f_{Ag} + f_{QD} + f_h = 1.
\end{equation}

The relative permittivity of the silver nanorods is expressed as a
function of angular frequency $\omega$  by the Drude formula \c{BH}
\begin{equation}
\eps_{Ag} (\omega) = 1 - \frac{\omega^2_p}{\omega^2 + i \, \omega \,
\gamma_{Ag} },
\end{equation}
where the plasma frequency $\omega_p = 1.38 \times 10^{16}$ rad
s${}^{-1}$ and the relaxation rate $\gamma_{Ag} = 10^{14}$
s${}^{-1}$. The relative permittivity of the quantum dots is
provided as \c{Bratkovsky}
\begin{equation}
\eps_{QD} (\omega) = \eps_b + \frac{a}{\omega - \omegao + i \,
\gamma_{QD} }.
\end{equation}
Herein, $\hbar \omega  = 0.8$ eV,  $\hbar a = 22.9 $ meV, where
$\hbar$ is the reduced Planck constant,  the background relative
permittivity $\eps_b = 11.8$, and the broadening parameter
$\gamma_{QD} = a / 15.3$ \c{Holm}. The host dielectric material is
taken to have the frequency--independent relative permittivity $\eps_h = 3$. The
real and imaginary parts of $\eps_{Ag}$ and $\eps_{QD}$ are plotted
against free--space wavelength $\lambda_o \in \les 1, 2 \ris$ $\mu$m
in Fig.~\ref{Fig1}. We see that  $\mbox{Im} \, \eps_{QD} < 0$ which
indicates the quantum dots are active entities, particularly so at
the resonance  centred on $\lambda_o = 1.55 $ $\mu$m.
 From the point of view of the applicability of
homogenization formalisms, there are two key points to notice: (i)
the quantity
\begin{equation}
\delta = \frac{\mbox{Re} \, \eps_{Ag}}{\mbox{Re} \, \eps_{QD}}
\end{equation}
is negative  over the entire wavelength range considered; and
(ii)  we have $\left| \, \mbox{Re} \, \eps_{Ag} \, \right| \gg
\left| \, \mbox{Im} \, \eps_{Ag} \, \right|$ for $\lambda_o \in \les
1, 2 \ris$ $\mu$m, and $\left| \, \mbox{Re} \, \eps_{QD} \, \right|
\gg \left| \, \mbox{Im} \, \eps_{QD} \, \right|$ for $\lambda_o \in
\les 1, 1.55 - \eta \ris \cup \les 1.55 + \eta, 2 \ris $ $\mu$m
where $ \eta$ specifies a sharp resonance region centred on $\lambda_o
= 1.55 $ $\mu$m. By analogy with dissipative  HCMs
\c{ML04,M07,DML07}, points (i) and (ii) suggest that  the
application of homogenization formalisms may be problematic.

\subsection{Homogenization formalisms}

The HCM is
characterized by the relative permittivity dyadic
\begin{equation}
\=\eps^{HCM} =  \eps^{HCM}_t  \, \le \, \=I - \hat{\#z} \, \hat{\#z}
\ri +  \eps^{HCM}_z \, \hat{\#z} \, \hat{\#z}.
\end{equation}
We write `MG' or `Br' in lieu of `HCM' according to whether the
Maxwell Garnett or the Bruggeman estimate of $\=\eps^{HCM}$ is being
considered. The anisotropy of the HCM arises due to the orientation
of  silver nanorods.

The Maxwell Garnett
 estimate of the relative permittivity
dyadic is given explicitly by \c{ML_PiO}
\begin{eqnarray}
 \=\eps^{MG} = \eps_h \=I + f_{Ag} \,\=a^{Ag/h} \. \le \=I -
 \frac{f_{Ag}}{3 \eps_h}
\,   \=a^{Ag/h} \ri^{-1} +  f_{QD} \, \=a^{QD/h} \. \le \=I -
\frac{f_{QD}}{3 \eps_h} \,  \=a^{QD/h} \ri^{-1},
\end{eqnarray}
wherein the polarizability density dyadics
\begin{equation} \l{pd_def}
\=a^{\ell/h} = \le \eps_{\ell} - \eps_h \ri \les \, \=I + \le
\eps_{\ell} - \eps_h \ri \=D^{\ell/h} \ris^{-1}, \qquad (\ell = Ag,
QD),
\end{equation}
and the depolarization dyadics \c{WM_02}
\begin{equation} \l{depol_def}
\left.
\begin{array}{l}
\=D^{Ag/h} = \displaystyle{ \frac{1}{2 \eps_h} \, \le \, \=I -
\hat{\#z} \, \hat{\#z} \ri} \vspace{4pt}
\\
\=D^{QD/h} = \displaystyle{\frac{1}{3 \eps_h} \,\=I}
\end{array}
\right\}.
\end{equation}

The Bruggeman  estimate
of the relative permittivity dyadic is given implicitly by
\c{ML_PiO}
\begin{eqnarray} \l{Br_eqn}
  f_{Ag} \,\=a^{Ag/Br} +
f_{QD} \,\=a^{QD/Br} + f_{h} \,\=a^{h/Br} = \=0 \,,
\end{eqnarray}
wherein the polarizability density dyadics
\begin{equation} \l{pd_def2}
\=a^{\ell/Br} = \le \eps_{\ell} \=I  - \=\eps^{Br} \ri \.  \les \,
\=I +  \=D^{\ell/Br} \. \le \eps_{\ell} \=I  - \=\eps^{Br} \ri
\ris^{-1}, \qquad (\ell = Ag, QD, h),
\end{equation}
and the depolarization dyadics \c{WM_02,Michel}
\begin{equation} \l{depol_def2}
\left.
\begin{array}{lcr}
\=D^{Ag/Br} = \displaystyle{ \frac{1}{2 \eps_t^{Br}} \, \le \, \=I -
\hat{\#z} \, \hat{\#z} \ri  } && \vspace{4pt}
\\
\=D^{\ell/Br} = \displaystyle{\frac{1}{ \eps^{Br}_t} \, \les \,
\alpha L (\alpha) \le \, \=I - \hat{\#z} \, \hat{\#z} \ri + L_z
(\alpha) \, \hat{\#z} \, \hat{\#z} \ris}\,, && (\ell = QD,h)
\end{array}
\right\}.
\end{equation}
Herein, the scalar functions
\begin{equation}
\left.
\begin{array}{l}
L(\alpha) = \displaystyle{\frac{1}{2 \alpha} \lec F(\alpha) -
\frac{1}{\alpha - 1} \les \, 1 - F (\alpha) \ris \ric} \vspace{4pt}
\\
L_z (\alpha) = \displaystyle{ \frac{1}{\alpha - 1} \les \, 1 - F
(\alpha) \ris }
\end{array}
\right\},
\end{equation}
with
\begin{equation}
F(\alpha) = \frac{1}{\sqrt{\alpha - 1}} \, \tan^{-1} \le
\sqrt{\alpha - 1} \ri
\end{equation}
and the scalar $\alpha = \eps^{Br}_z / \eps^{Br}_t$.

The relative permittivity dyadic $\=\eps^{Br}$ may be extracted from
the nonlinear equation \r{Br_eqn} by numerically. For example, the
 iterative scheme \c{ML06,Michel_chap}
\begin{equation}
\=\eps^{Br}_{\,(j)} = \mathcal{T} \lec \=\eps^{Br}_{\,(j-1)} \ric,
\qquad \qquad \le \, j = 1, 2, 3, \ldots \, \ri,
\end{equation}
wherein the operator $\mathcal{T}$ is defined as
\begin{eqnarray}
\mathcal{T} \lec \=\eps^{Br} \ric &=& \Big\{ f_{Ag} \, \eps_{Ag}
\les \, \=I + \=D^{Ag/Br} \. \le \eps_{Ag} \=I - \=\eps^{Br} \ri
\ris^{-1} + f_{QD} \, \eps_{QD} \les \, \=I + \=D^{QD/Br} \. \le
\eps_{QD} \=I - \=\eps^{Br} \ri \ris^{-1}  \nonumber \\ && + f_{h}
\, \eps_{h} \les \, \=I + \=D^{h/Br} \. \le \eps_{h} \=I -
\=\eps^{Br} \ri \ris^{-1} \Big\} \. \Big\{ f_{Ag}  \les \, \=I +
\=D^{Ag/Br} \. \le \eps_{Ag} \=I - \=\eps^{Br} \ri \ris^{-1}
\nonumber \\ && f_{QD}  \les \, \=I + \=D^{QD/Br} \. \le \eps_{QD}
\=I - \=\eps^{Br} \ri \ris^{-1} + f_{h}  \les \, \=I + \=D^{h/Br} \.
\le \eps_{h} \=I - \=\eps^{Br} \ri \ris^{-1}  \Big\}^{-1},
\end{eqnarray}
and the initial value
\begin{equation}
\=\eps^{Br}_{\,(0)} = \le \, f_{Ag} \eps_{Ag} + f_{QD} \eps_{QD} +
f_{h} \eps_h \, \ri \, \=I\,,
\end{equation}
works sufficiently well for our purposes here.

\subsection{Numerical estimates of the HCM permittivity}

Let us begin by considering the Maxwell Garnett and Bruggeman
estimates for an active HCM when the parameter $\delta$ is positive.
This is achieved by setting $f_{Ag} = 0$. Thus, the composite
material comprises only  quantum dots and the host dielectric
material, and the resulting HCM  is an isotropic dielectric material
with relative permittivity dyadic $\eps^{HCM}\=I$. Estimates of the
HCM relative permittivity computed using the Maxwell Garnett
formalism and the Bruggeman formalism, namely $\eps^{MG}$ and
$\eps^{Br}$ respectively, are plotted against $\lambda_o$ in
Fig.~\ref{Fig2}, when $f_{QD} = 0.3$. The Maxwell Garnett and
Bruggeman estimates are similar, both qualitatively and
quantitatively. For both formalisms, the HCM is active across the
entire wavelength range considered, and its permittivity scalar
exhibits a sharp resonance at $\lambda_o = 1.55$ $\mu$m. Therefore,
we infer that both the Maxwell Garnett and the Bruggeman formalisms
are suitable for active HCMs with $\delta > 0$.

Next we turn to the $\delta < 0$ regime which is known to be
problematic for dissipative  HCMs \c{ML04,M07,DML07}. In keeping
with Bratkovsky \emph{et al.} \c{Bratkovsky}, we fix $f_{Ag} =
0.063$ and $f_{QD} = 0.3$; thus, $f_h$ is slightly reduced from its
value for Fig.~\ref{Fig2}. In Fig.~\ref{Br_MG_Ag_fig}, the two
entries in of the  HCM's relative permittivity dyadic, as computed
using the Maxwell Garnett and Bruggeman formalisms, are plotted
against $\lambda_o$. While the Maxwell Garnett and Bruggeman graphs
in Fig.~\ref{Br_MG_Ag_fig} are qualitatively similar, there are
substantial quantitative differences. In particular, the imaginary
part of $\eps^{Br}_z$ is  more than 20 times larger than
$\eps^{MG}_z$, apart from at the small resonance region centred on
$\lambda_o = 1.55$ $\mu$m. There are also substantial differences
between Re $\eps^{Br}_z$ and Re $\eps^{MG}_z$. By comparison, the
differences between $\eps^{Br}_t$ and $\eps^{MG}_t$ are relatively
small.

For dissipative  HCMs, the problematic nature of the $\delta < 0$
regime is exacerbated as the imaginary parts of the constitutive
parameters tend towards zero  \c{ML04,M07,DML07}. In order to
investigate this issue for active HCMs, we repeated the computations
of Fig.~\ref{Br_MG_Ag_fig} with $\gamma_{Ag}$ and $\gamma_{QD}$ multiplied by
$10^{-6}$. The corresponding plots of the real and
imaginary parts of $\eps^{Br,MG}_{t,z}$ are presented in
Fig.~\ref{Br_MG_Ag_fig2}. The differences between the Maxwell
Garnett estimates and the Bruggeman estimates are now enormous,
especially between the imaginary parts of $\eps^{Br}_z$ and $\eps^{MG}_z$.

\section{Concluding remarks}

Our numerical studies reveal that the well--known homogenization
formalisms named after Maxwell Garnett and Bruggeman appear to be
fairly consistent, and provide physically plausible estimates, when
applied to active isotropic dielectric HCMs, provided that the real
parts of the relative permittivities of the component materials have the same
sign. In contrast, if the  real parts of the permittivities of the
component materials have different signs then there are substantial
differences between the  estimates yielded by the two formalisms.
These differences become enormous in the limiting case in which the
imaginary parts of the relative permittivities of the component materials
become vanishingly small.

 For the particular homogenization scenario considered here, in the $\delta < 0$ with
$\gamma_{Ag,QD} \to 0$ regime,  the HCM permittivity parameters
estimated using the Bruggeman formalism have relatively large
positive--valued imaginary parts away from the resonance region
centred on $\lambda_o = 1.55$ $\mu$m, and relatively large
negative--valued imaginary parts at the resonance region centred on
$\lambda_o = 1.55$ $\mu$m. This is not physically plausible since it
implies that the HCM is a strongly active material at the resonance
region, and a strongly dissipative material away from  the resonance
region, in the limit wherein the component materials become inactive
and nondissipative. Let us note that Bruggeman formalism arises as
the lowest--order formulation of the
strong--permittivity--fluctuation theory (SPFT)\footnote{Otherwise
known as the strong--property--fluctuation theory for more general
HCMs \c{MLW00}.} \c{TK81}. Accordingly, higher--order
implementations of the SPFT are also subject to the limitations of
the Bruggeman formalism. Furthermore, since the Maxwell Garnett and
Bruggeman formalisms share a common provenance \c{Aspnes}, doubt is
also cast over the applicability of the Maxwell Garnett formalism
for active HCMs with $\delta <0$ where the imaginary parts of the
relative permittivities of the component materials are relatively
small.

The present study extends and reinforces our previous studies which
highlighted the limitations of conventional homogenization
formalisms when applied to scenarios wherein the real parts of the
constitutive parameters characterizing  the component materials have
different signs \c{ML04,M07,DML07}. We see here that caution is
needed for active HCMs as well as dissipative  ones.

\vspace{10mm}

\noindent {\bf Acknowledgement:} TGM thanks Dr Petter Holmstr\"{o}m
(KTH--Royal Institute of Technology, Sweden) for his helpful
comments.

 \newpage

\begin{figure}[!ht]
\centering
\includegraphics[width=3.0in]{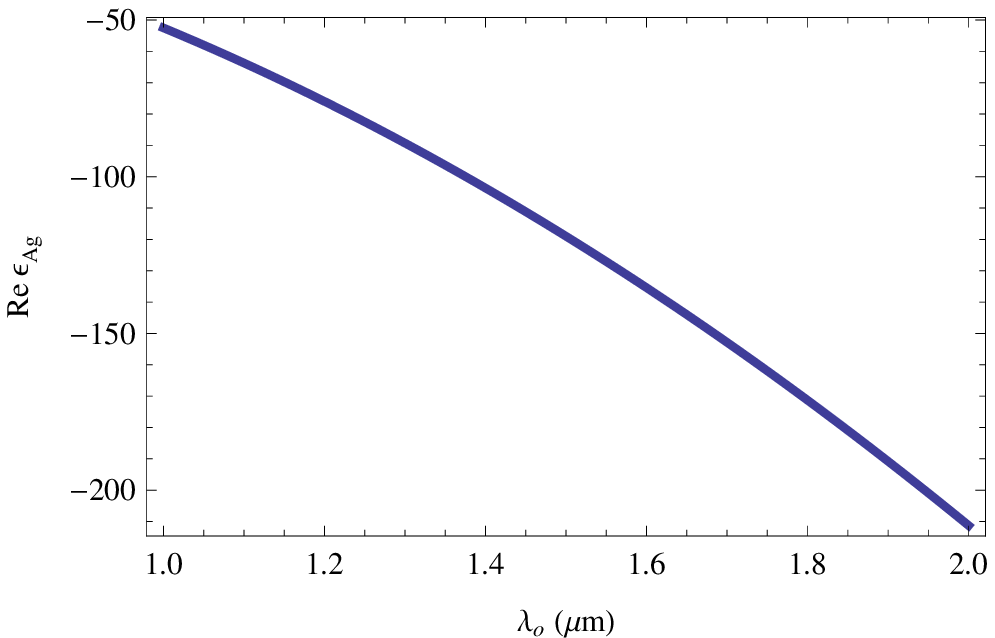}
 \includegraphics[width=3.0in]{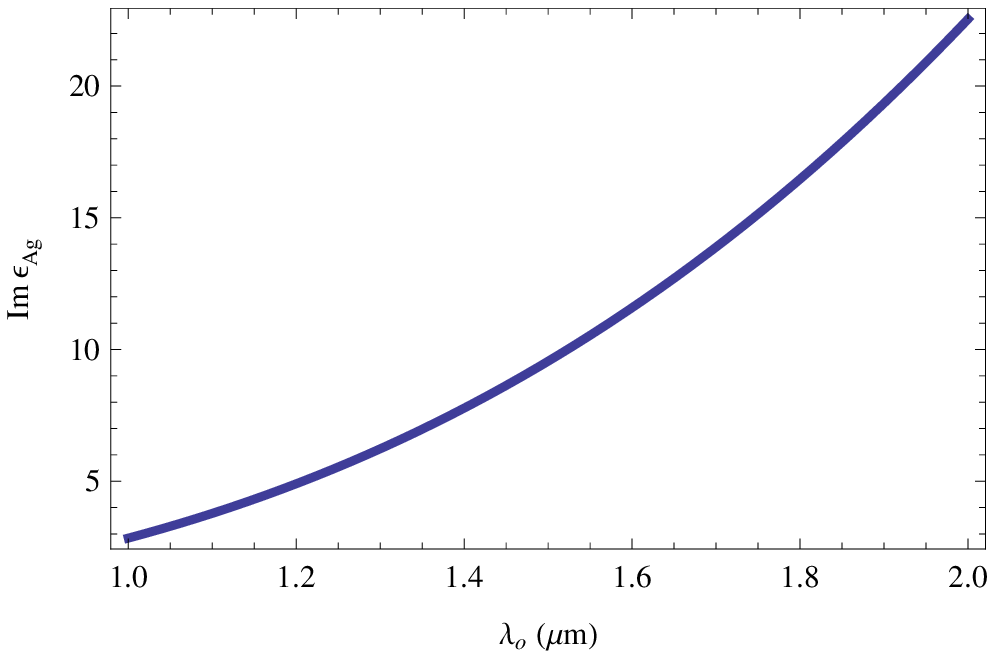}\\
\includegraphics[width=3.0in]{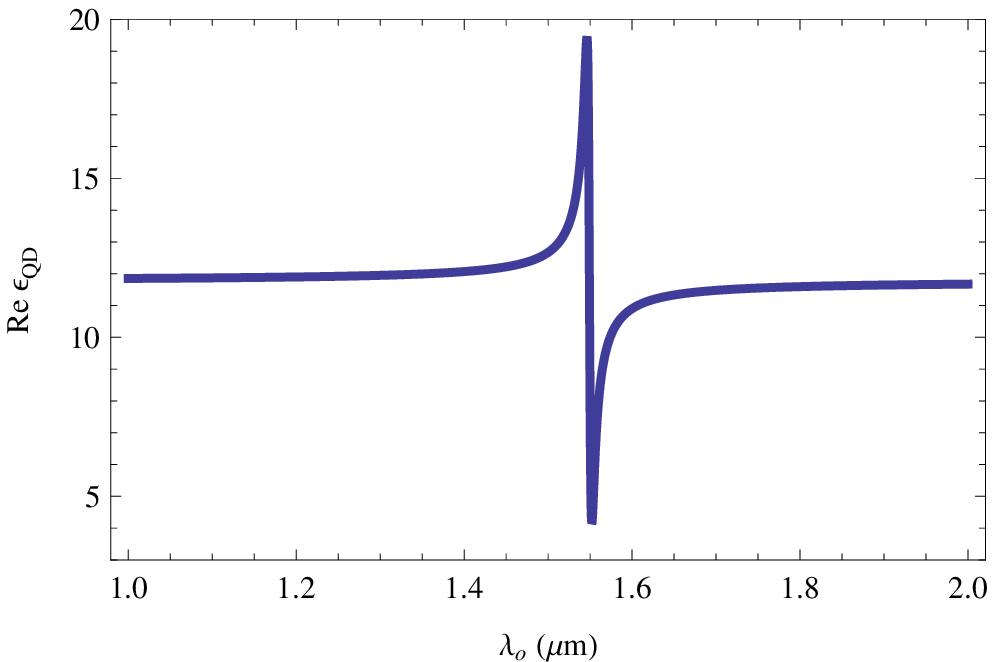}
 \includegraphics[width=3.0in]{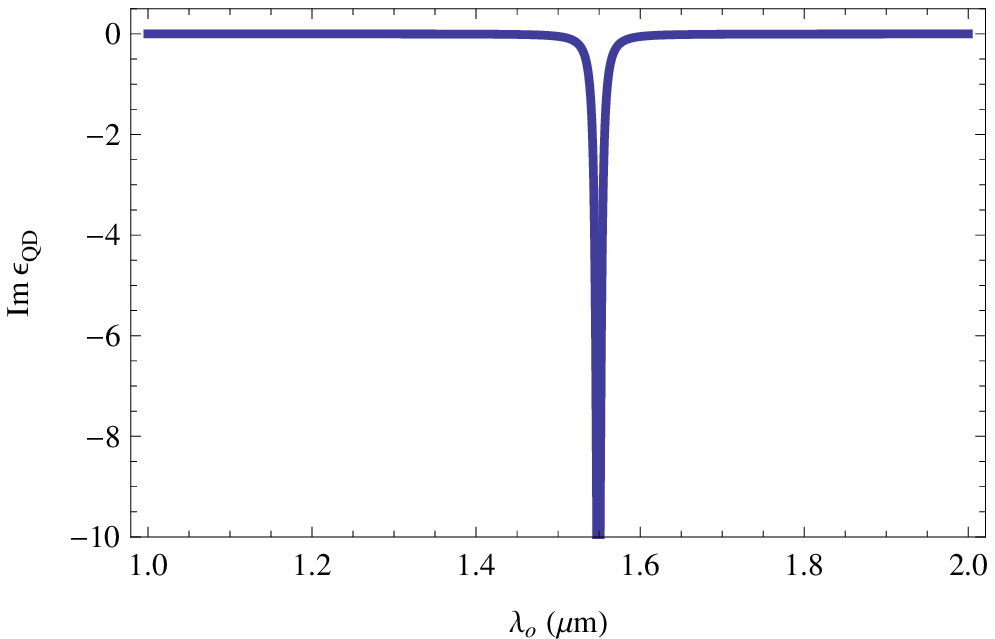}
 \caption{ \l{Fig1} Real and imaginary parts of the relative permittivities
 $\epsilon_{Ag}$ and $\epsilon_{QD}$
plotted against free--space wavelength $\lambda_o$ ($\mu$m).}
\end{figure}

\newpage
\vspace{10mm}

\begin{figure}[!ht]
\centering
\includegraphics[width=3.0in]{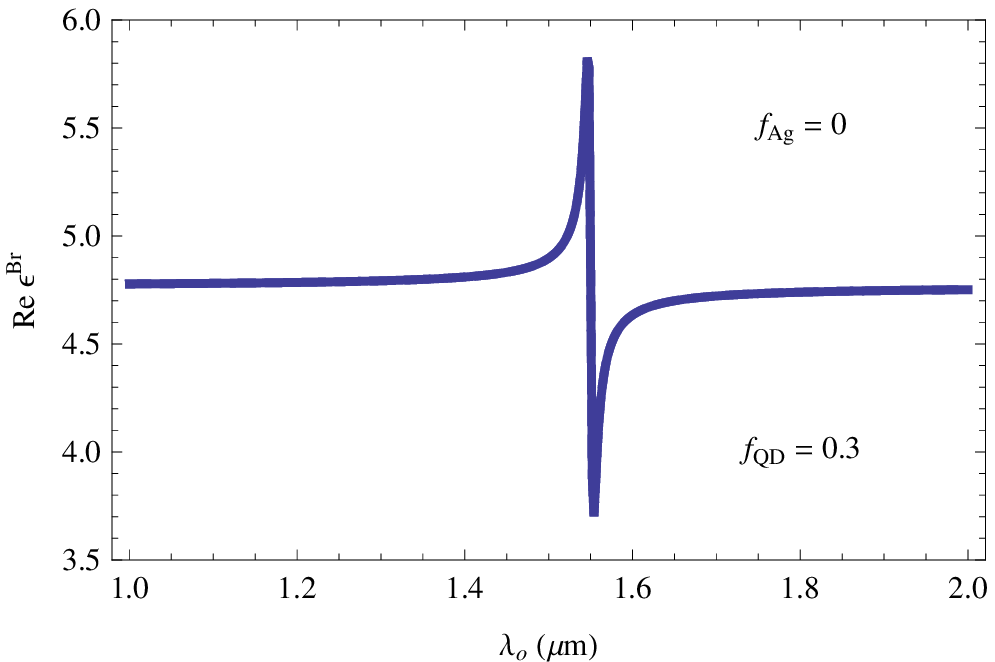}
 \includegraphics[width=3.0in]{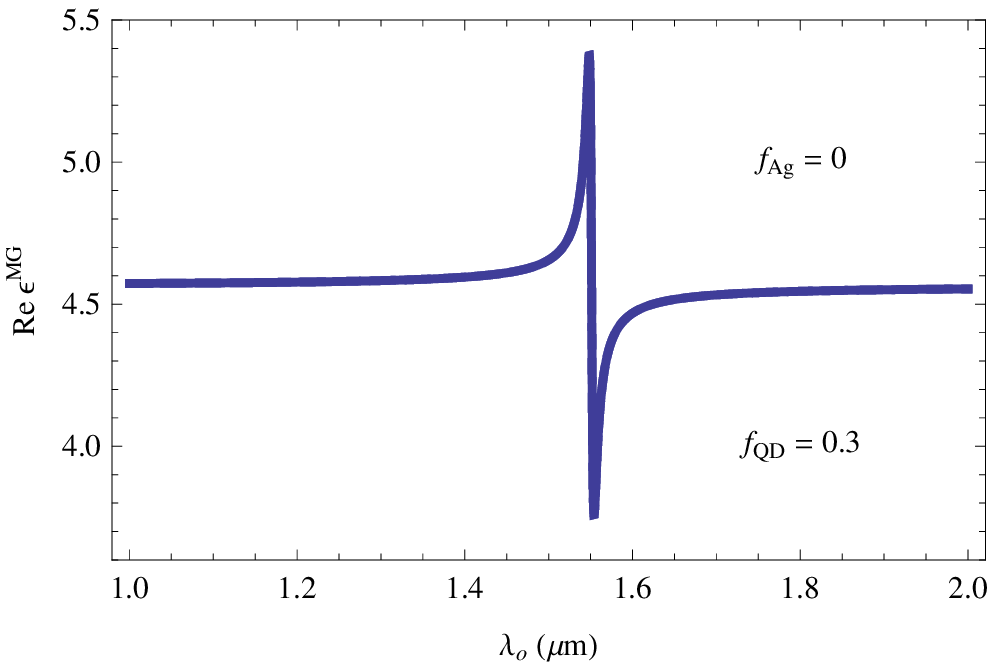}\\
\includegraphics[width=3.0in]{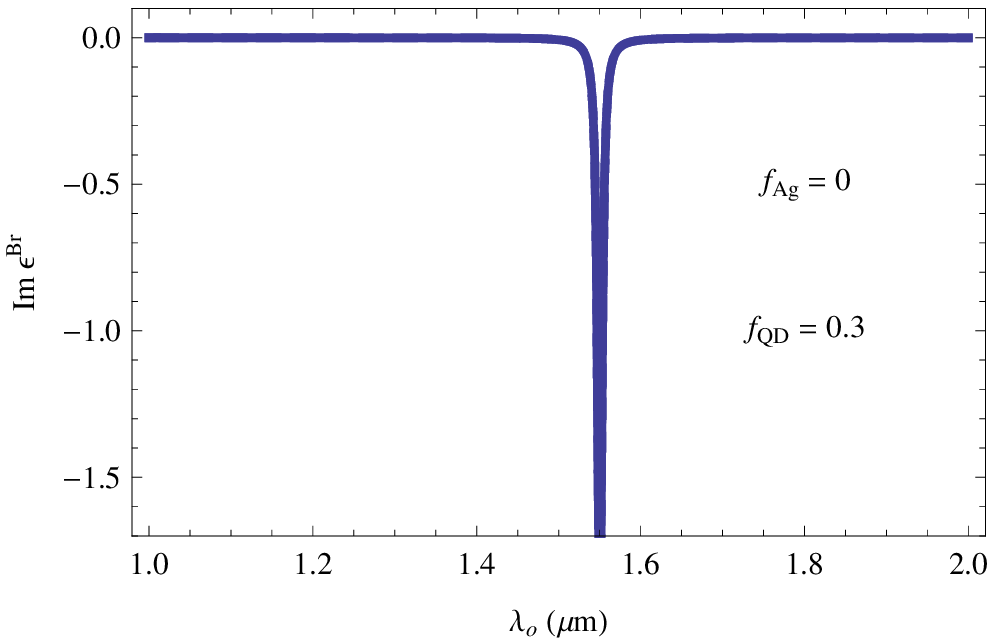}
 \includegraphics[width=3.0in]{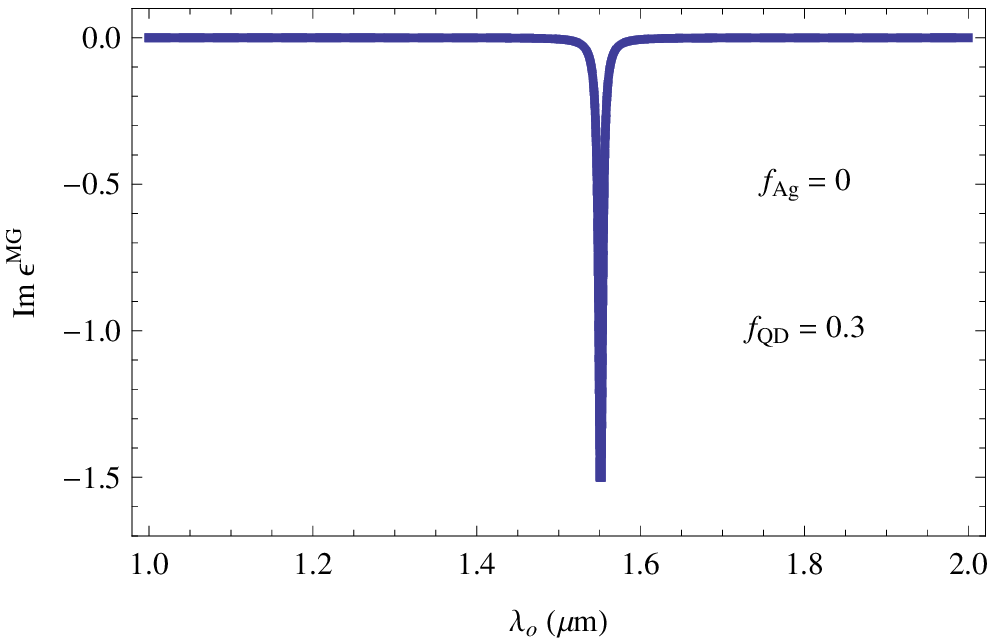}
 \caption{ \l{Fig2} Bruggeman (left) and Maxwell Garnett (right) estimates of
the HCM's relative permittivity  plotted against
 $\lambda_o$ ($\mu$m), with
  $f_{Ag} = 0$ and $f_{QD} = 0.3$.}
\end{figure}

\newpage
\vspace{10mm}

\begin{figure}[!ht]
\centering
\includegraphics[width=3.0in]{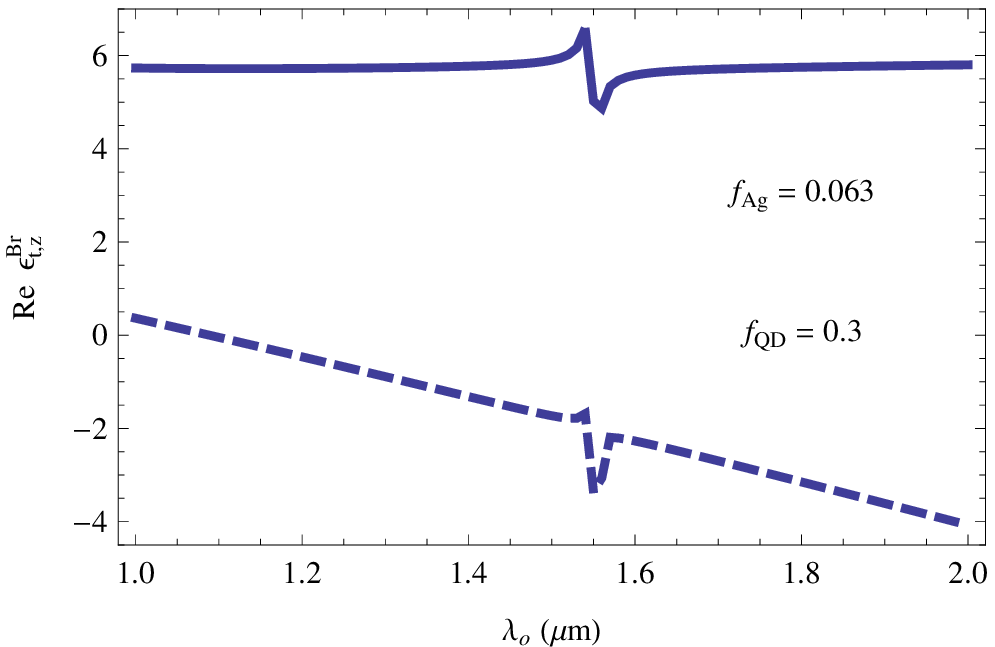}
 \includegraphics[width=3.0in]{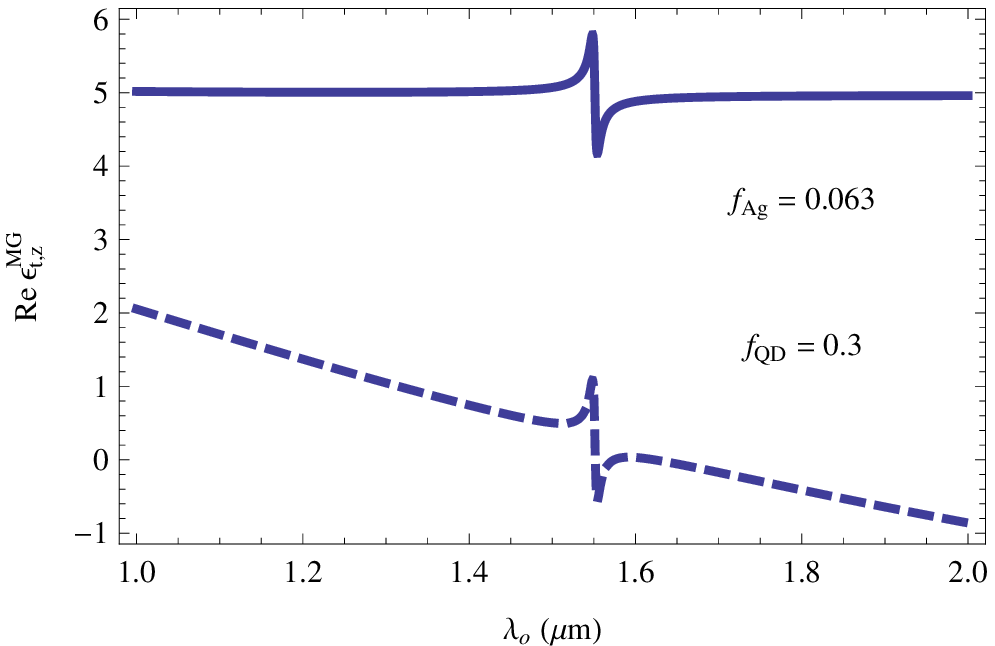}\\
\includegraphics[width=3.0in]{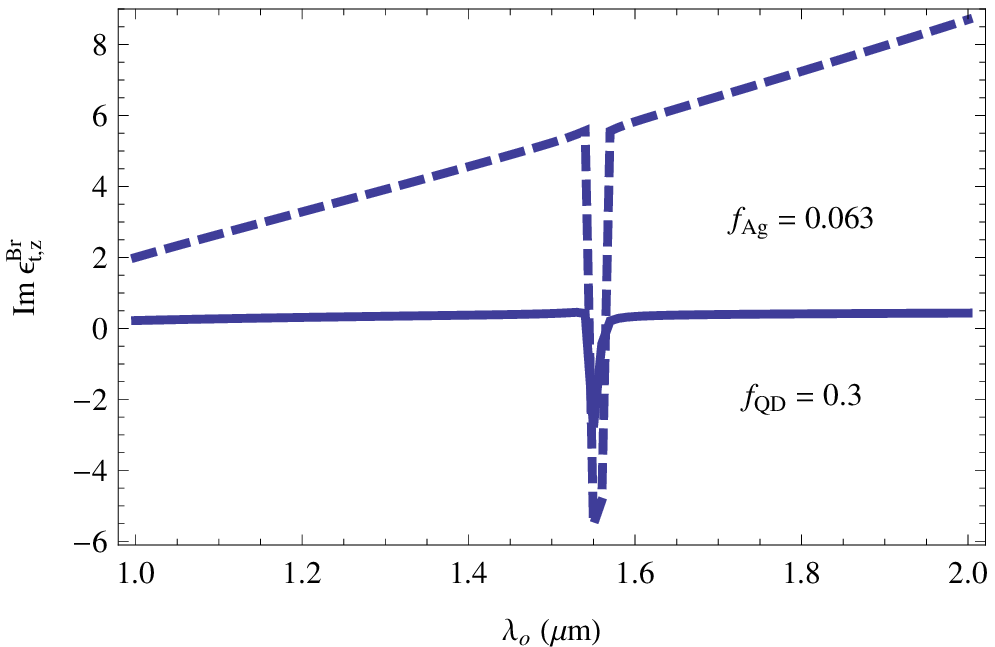}
 \includegraphics[width=3.0in]{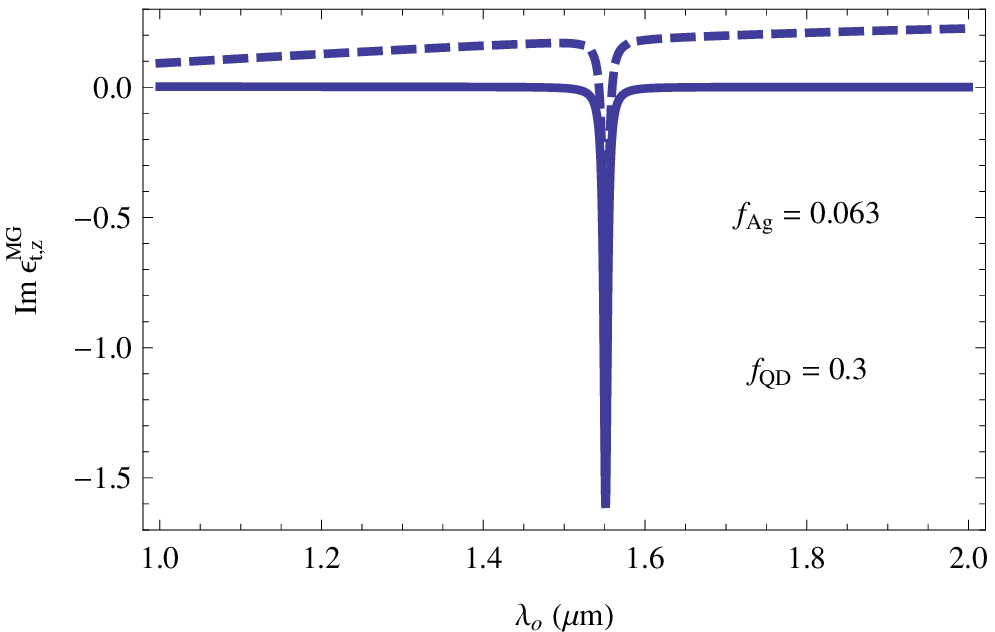}
 \caption{Bruggeman (left) and Maxwell Garnett (right) estimates of the two entries in
the HCM's relative permittivity dyadic plotted against
$\lambda_o$ ($\mu$m), with
  $f_{Ag} = 0.063$ and $f_{QD} = 0.3$. The solid curves represent $\eps^{Br,MG}_t$ whereas the dashed
  curves represent $\eps^{Br,MG}_z$. \label{Br_MG_Ag_fig}}
\end{figure}

\newpage

\vspace{10mm}

\begin{figure}[!ht]
\centering
\includegraphics[width=3.0in]{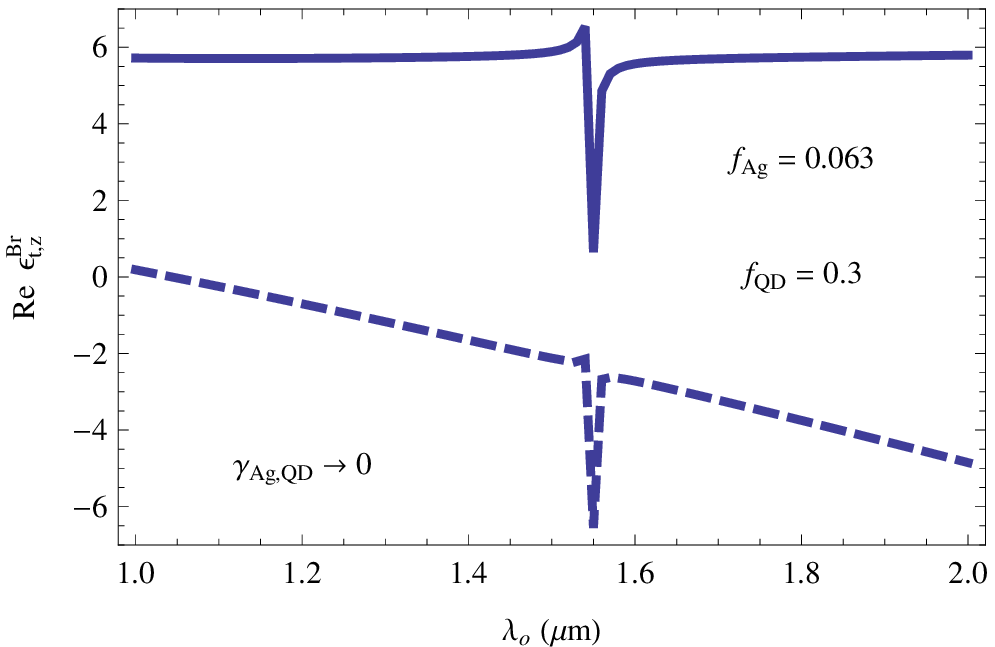}
 \includegraphics[width=3.0in]{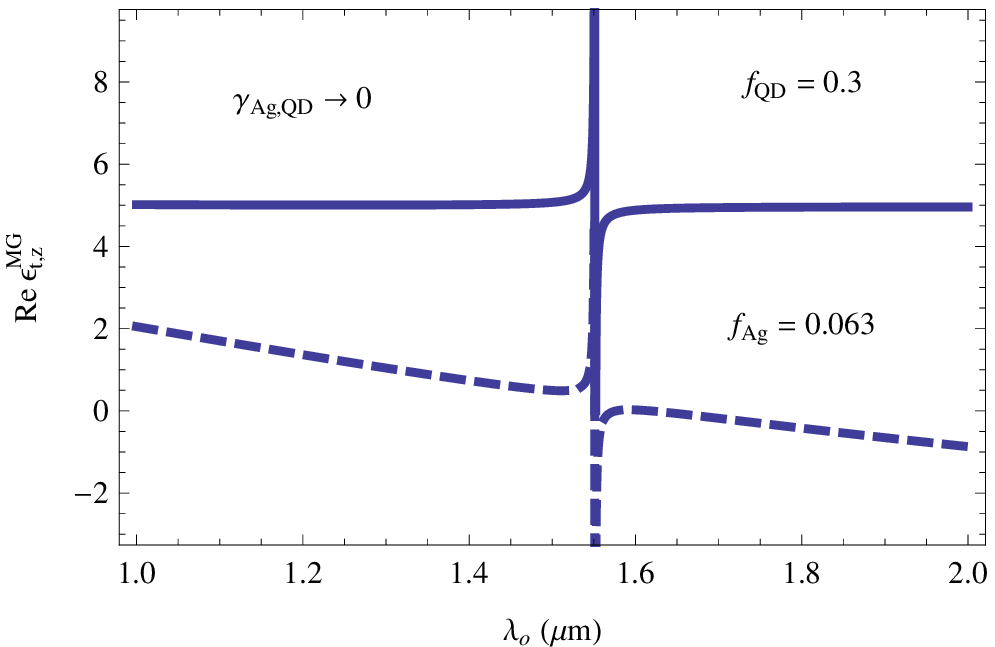}\\
\includegraphics[width=3.0in]{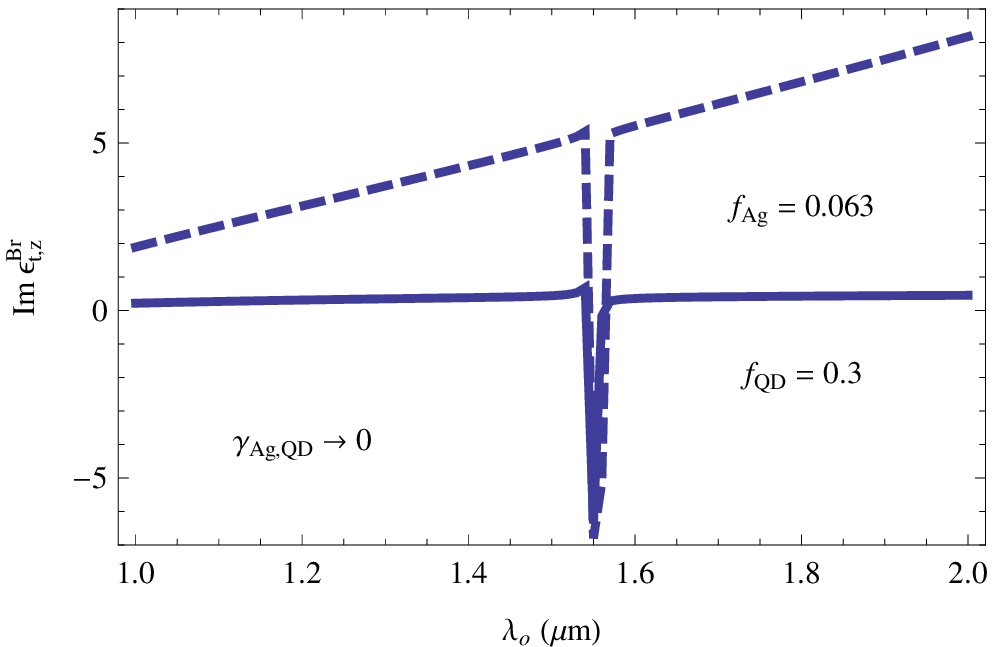}
 \includegraphics[width=3.0in]{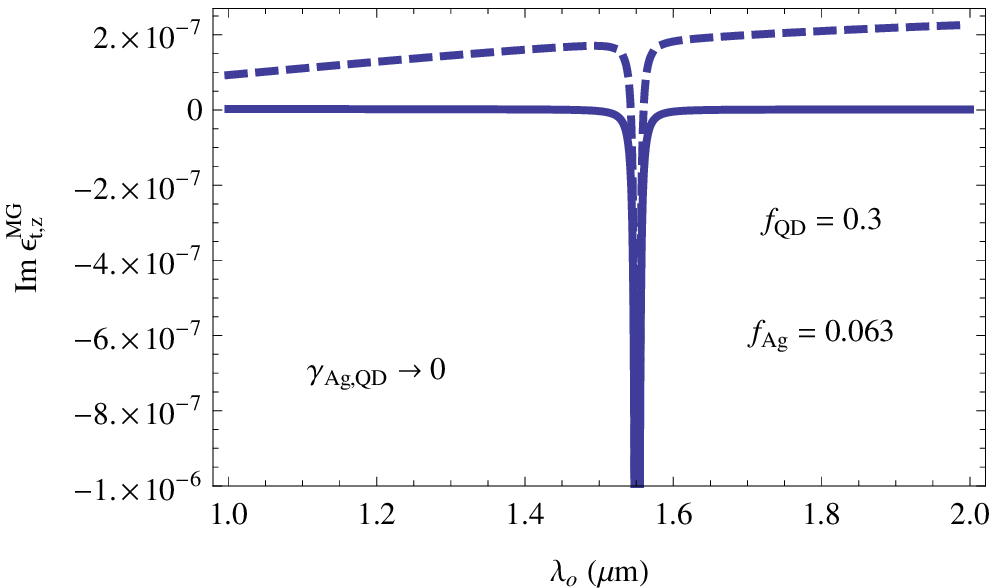}
 \caption{As Fig.~\ref{Br_MG_Ag_fig} but with  both $\gamma_{Ag}$ and  $\gamma_{QD}$ multiplied by a factor of $10^{-6}$.
 \label{Br_MG_Ag_fig2}}
\end{figure}


\begin{thebibliography}{99}



\bibitem{M08}
 T. G. Mackay,
J. Nanophoton.  2 (2008) 029503.


\bibitem{MG}
J. C. Maxwell Garnett, 
Phil. Trans. R. Soc. Lond. A  203 (1904)  385.
 (Reproduced in \c{L96}).

\bibitem{Smith}
G. B. Smith, J. Phys. D: Appl. Phys. 10 (1977) L39.


\bibitem{Niklasson}
G. A. Niklasson, C. G. Granqvist, O. Hunderi, Appl. Opt. 20 (1981)
26.

\bibitem{Brugge}
D. A. G. Bruggeman,
Ann. Phys. Lpz. 24 (1935) 636. (Reproduced in \c{L96}).

\bibitem{Goncharenko}
A. V. Goncharenko, Phys. Rev. E 68 (2003) 041108. Corrections: 69
(2004) 029905.

\bibitem{ML06}
 T. G. Mackay,  A. Lakhtakia, 
Opt. Commun.  259 (2006) 727.


\bibitem{L96}
A. Lakhtakia (Ed.),  Selected Papers on Linear Optical Composite
Materials, SPIE, Bellingham, WA, USA, 1996.

\bibitem{ML04}
T. G. Mackay,  A. Lakhtakia,
Opt. Commun.  234 (2004) 35.


\bibitem{M07}
T. G. Mackay, 
  J. Nanophoton.  1 (2007)  019501.

\bibitem{DML07}
A. J. Duncan, T. G. Mackay,  A. Lakhtakia, 
Opt. Commun.  271 (2007) 470.


\bibitem{M05}
T. G. Mackay,
{Electromagnetics} { 25} (2005) 461.

\bibitem{Noginov}
M. A. Noginov, 
J. Nanophoton. 2 (2008) 021855.

\bibitem{Bratkovsky}
A. Bratkovsky, E. Ponizovskaya, S.--Y. Wang, P. Holmstr\"{o}m, L.
Thyl\'{e}n, Y. Fu, H. ${\buildrel _{\circ} \over {\mathrm{A}}}$gren,
Appl. Phys. Lett.  93 (2008) 193106.


\bibitem{BH}
C. F.  Bohren,  D. R.  Huffman,  Absorption and Scattering of Light
by Small Particles, Wiley, New York, NY, USA, 1983.

\bibitem{Holm}
P. Holmstr\"{o}m. Personal communication (December 2008). Note that
the parameter values of $a$ and $\gamma_{QD}$ we adopt here differ
slightly from used for the computations described by Bratkovsky
\emph{et al.} \c{Bratkovsky} (who accidentally used a gain value
which is too low).


\bibitem{ML_PiO}
T. G. Mackay, A. Lakhtakia,
  Prog. Optics  51 (2008)  121.

\bibitem{WM_02}
W. S. Weiglhofer,  T. G. Mackay, 
IEEE Trans.  Antennas  Propagation.  50 (2002) 85.

\bibitem{Michel}
B. Michel, 
Int. J. Appl. Electromagn. Mech.  8 (1997) 219.

\bibitem{Michel_chap}
B. Michel,
  in:  O.N. Singh, A. Lakhtakia (Eds.),
Electromagnetic Fields in Unconventional Materials and Structures,
Wiley, New York, NY, USA, 2000, p. 39.

 \bibitem{MLW00}
T. G.  Mackay, A.   Lakhtakia, W. S.   Weiglhofer,
Phys. Rev. E  62 (2000) 6052.  
Corrections:   63 (2001) 049901.


\bibitem{TK81}
L.  Tsang, J. A.  Kong, 
Radio Sci.
16 (1981) 303. 
 (Reproduced in \c{L96}).

\bibitem{Aspnes}
D. E. Aspnes, 
Am. J. Phys.  50 (1982)  704.  
  (Reproduced in \c{L96}).


\end{thebibliography}
\end{document}